\documentstyle[12pt,twoside,fleqn,espcrc1,epsfig]{article}

\newcommand{\AmS}{{\protect\the\textfont2
  A\kern-.1667em\lower.5ex\hbox{M}\kern-.125emS}}
\input{epsf}
\begin{document}
\title{Resonance studies at STAR}
\author{
Zhangbu Xu, Yale University, for The STAR Collaboration$^{\cite{harris}}$}
\maketitle
\vspace{-0.3cm}
\begin{abstract}
We report on the observed signals of 
${K^{\star0}(892)}\rightarrow\pi K$ and 
$\phi(1020)\rightarrow K^{+}K^{-}$ 
using the mixed-event method with powerful statistics from the 
large acceptance and highly efficient STAR TPC. 
Preliminary results from the first observation of such states from 
the year-one STAR data in $\sqrt{s_{NN}}=130$ GeV Au-Au collisions 
are presented. The $K^{\star0}/h^{-}$ ratios with an assumed $K^{\star0}$ 
$p_{T}$ inverse slope of 300MeV are compatible with that from pp at 
ISR. For 14\% central Au+Au collisions, we observe 
$K^{\star0}/h^{-}=0.060\pm0.007(stat)$ and 
$\overline{K^{\star0}}/h^{-}=0.058\pm0.007(stat)$.
We show that $\overline{\Lambda}/\Lambda=0.77\pm0.07(stat)$ from this method 
is consistent with the measurement via decay topology. 
\end{abstract}
\vspace{-0.4cm}
\section{Introduction}
\vspace{-0.4cm}
 Resonance (vector meson, etc.) production and the modification to their 
properties by the medium are important signals of
phase transition in relativistic heavy ion collisions. 
Leptonic decay channels of vector mesons 
have been extensively studied~\cite{muller}. 
Their dominant decay branches to hadrons, however, are less extensively 
studied, partially due to the
final state interactions of decay products, the large background 
from the abundantly produced hadrons ($\pi,K$'s) and the broad mass width.
This is especially true for mesons like $\rho(770)$ and $K^{\star0}(892)$ whose
daughters consist of $\pi$'s and the decay widths are 130MeV and 50MeV, 
respectively. 
On the other hand, due to their short lifetime which is comparable to the 
lifetime of the dense matter, their measured properties may be sensitive 
to the lifetime of the dense matter. For example, model calculations show 
that the $K^{\star0}/K$ ratio is sensitive to the mass 
modification~\cite{schaffner} of particles in the medium and the dynamic 
evolution of the source. From detailed comparison of yields and 
$p_{T}$ distributions of resonances and other particles, we 
may be able to distinguish different freeze-out 
conditions~\cite{rafelski,bravina}. 

Simulation has shown that we should be able to 
reconstruct these resonances with good statistics at RHIC energies through 
the mixed-event method because the significance of the signal rises with the 
square root of number of events~\cite{xzb}. 
However, other effects from flow and detector complicate the situation when 
the signal is less than 1\% of the combinatorics. We can also apply the same 
mixed-event methos to reconstruct lambda, l-bar, and Ks which are 
complementary to those from secondary vertex identification.
\vspace{-0.4cm}
\section{Experimental setup and data analysis}
\vspace{-0.4cm}
  The main detector for STAR is the Time Projection Chamber(TPC)~\cite{harris} 
allowing measurement of track multiplicity and momenta of the 
tracks. The Zero Degree Calorimeters (ZDC) are used to define a minimum bias 
trigger and an additional scintillating central trigger barrel (CTB) to select 
the top 14\% central events~\cite{harris}. 
In this analysis, we take advantage of STAR TPC's large acceptance 
($\approx95\%$) and high efficiency ($\approx85\%$) at mid-rapidity 
($|y|<0.5$) to overcome the large combinatoric background. 

Data were taken in the summer of 2000 for Au+Au collisions with 
$\sqrt{s_{NN}}=130$ GeV~\cite{flow}. There are 307K central and 160K 
minimum bias Au+Au nuclear interactions surviving the cuts described below 
and used in this analysis. 
Event vertex is required to be within $|Z|<95cm$ along the 
beam direction of the center of the TPC for uniform acceptance in the range we 
study~\cite{manuel}. Particles are selected based on momenta measured from 
track curvature in the TPC at 0.25T solenoidal magnetic 
field, track quality and particle identification from energy loss in the TPC. 
The cuts used depend on the daughter particle species. 
Since $K\pi$ from $K^{\star0}$ and $\phi$ decays are particles originating 
from the interaction point, we select tracks with $DCA<3$ cm where
DCA is the distance of closest approach to the primary vertex.
For $p,\overline{p},\pi$ from $\Lambda$, 
$\overline{\Lambda}$ with $c\tau=7.8cm$, we select tracks with $DCA<7$ cm. 
The purity of the PID selection is good only for low momentum. 
For example, for $p>0.8$ GeV/$c$, more than 80\% of the produced 
particles have dE/dx consistent with being a kaon. 
In addition, we require that pseudorapidity of the daughters $|\eta|<0.8$ 
and opening angle between the daughters $\Delta \theta>0.2$ for 
$K^{\star0}$ and $\Delta \theta>0.05$ for $\phi$ and $\Lambda$. 

\begin{figure}
\vspace{-1.cm}
\begin{minipage}[b]{.46\linewidth}
\centering
\epsfig{file=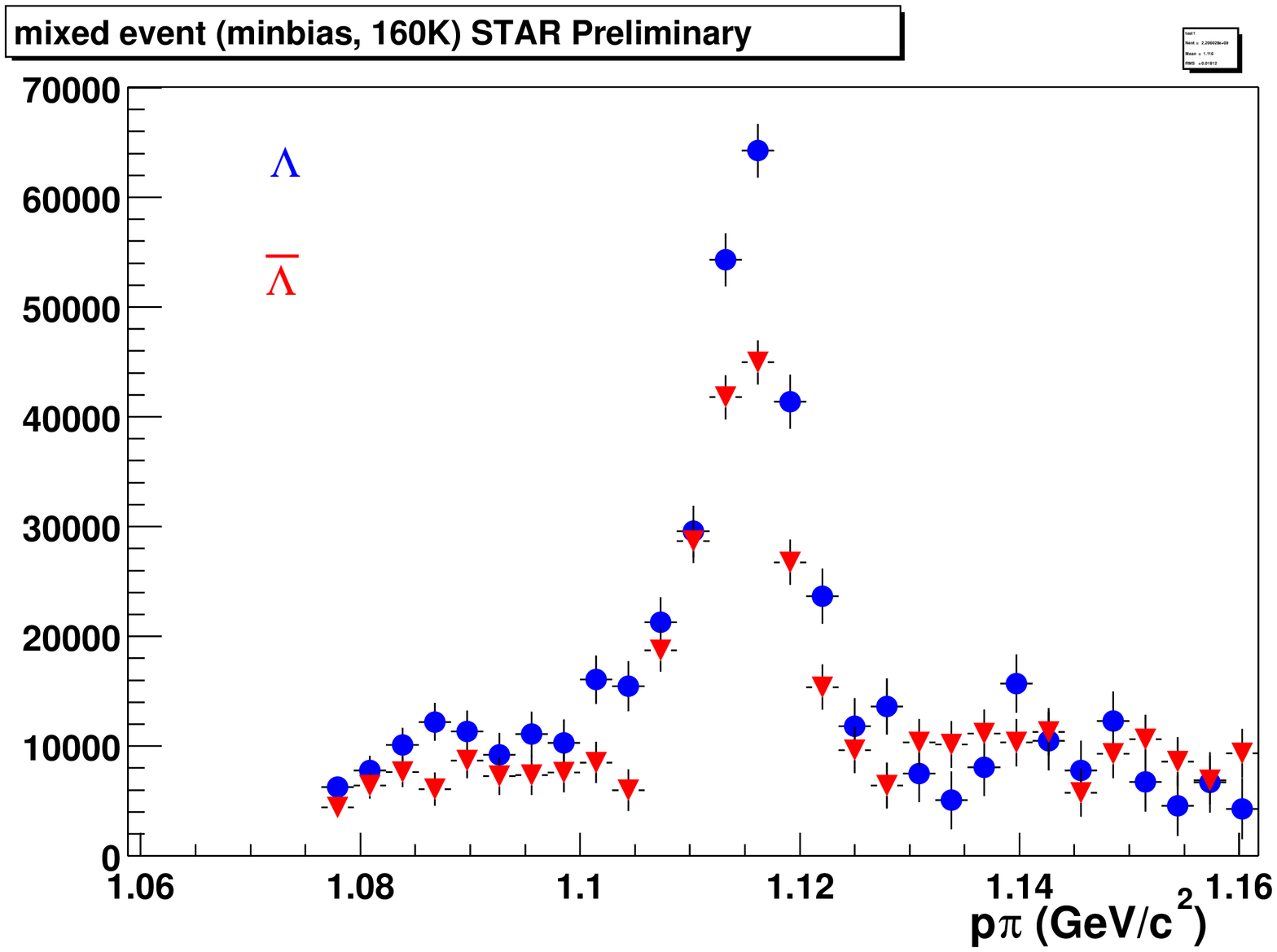,height=1.9in,angle=270}
\vspace{-0.45in}
\caption{\protect{$\Lambda$} and 
\protect{$\overline{\Lambda}$} mass plot after mixed-event background 
subtraction from 160K minbias events,
\protect{$\overline{\Lambda}/\Lambda=0.77\pm0.07(stat)$}.\vspace{-0.9cm}}
\label{fig:lambda}
\end{minipage}
\begin{minipage}[b]{.46\linewidth}
\centering\epsfig{file=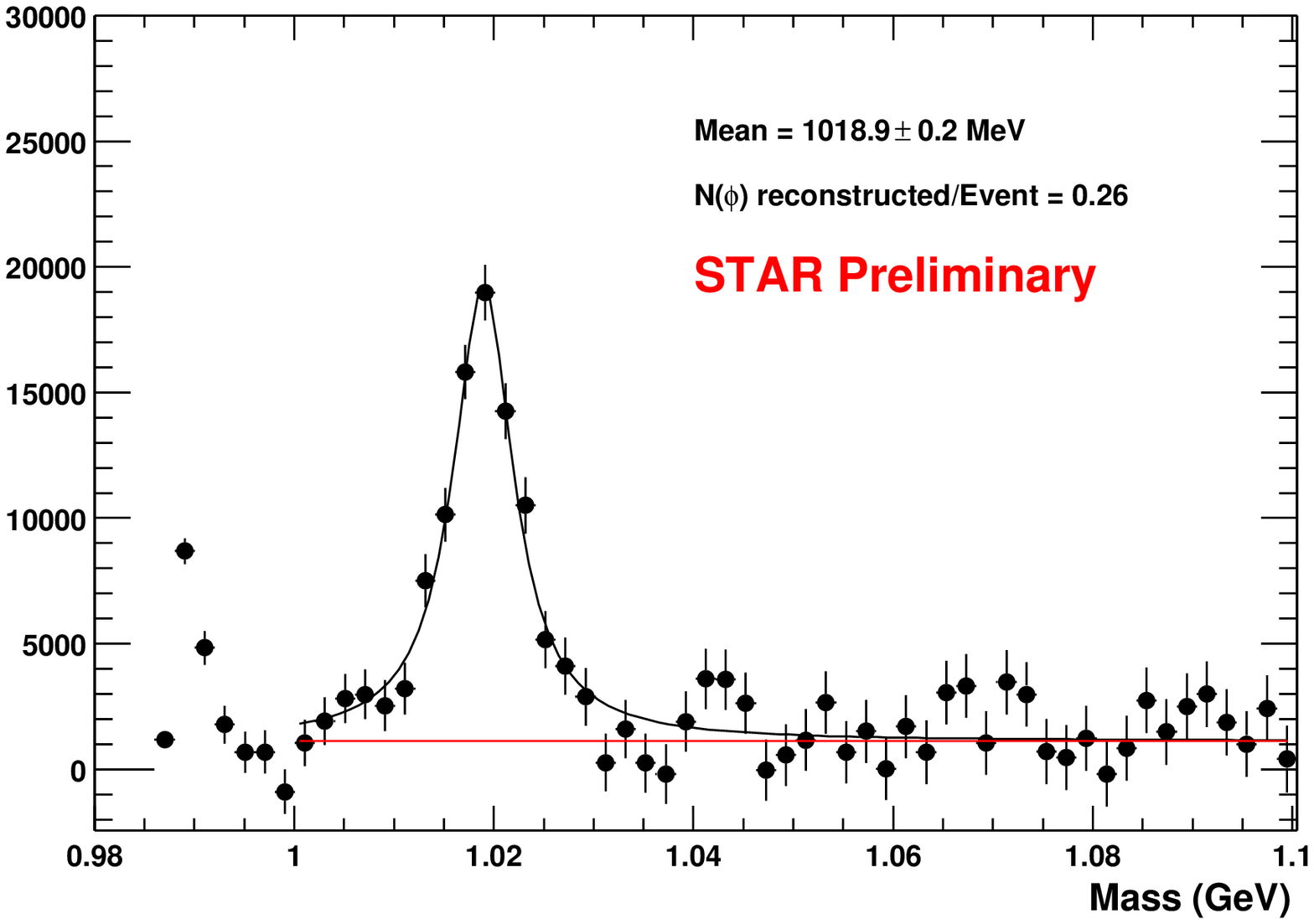,height=2.in}
\vspace{-0.45in}
\caption{\protect{$\phi$} mass plot after mixed-event background 
subtraction from 400K central event.\vspace{-0.9cm}
}
\label{fig:phi}
\end{minipage}
\hspace{\fill}
\end{figure}

Fig.\ref{fig:lambda} and Fig.\ref{fig:phi} shows the mixed-event subtracted 
mass plots of $\Lambda$, $\overline{\Lambda}$ and 
$\phi$. From these distributions, we fit the $\Lambda$ and 
$\overline{\Lambda}$ mass peaks and calculate the $p_{T}$ and rapidity 
integrated ratio of $\overline{\Lambda}/\Lambda=0.77\pm0.07(stat)$ for 
minimum-bias events. 
The $p_{T}$ and rapidity range  of the parent particle are $p_{T}<2$ GeV/$c$ 
and $|y|<0.5$ in addition to the cuts on daughter particles.  
In this calculation, the efficiencies of $\Lambda$ and $\overline{\Lambda}$ 
cancel out due to the cylindal symmetry of the TPC and its magnetic field. 
This result is consistent with the result of 
$\overline{\Lambda}/\Lambda=0.73\pm0.03(stat)$ from topological 
method~\cite{manuel}. 
It shows that there is net strange baryon in the middle rapidity. 
We observe about 0.26 reconstructed $\phi$ per central event with good 
statistics from $\phi\rightarrow K^{+}K^{-}$ (B.R. 49.1\%). 
Analysis on $p_{T}$ spectra and centrality dependence is under way. 

\begin{figure}[htb]
\vspace{-1cm}
\begin{minipage}[b]{.46\linewidth}
\centering
\epsfig{file=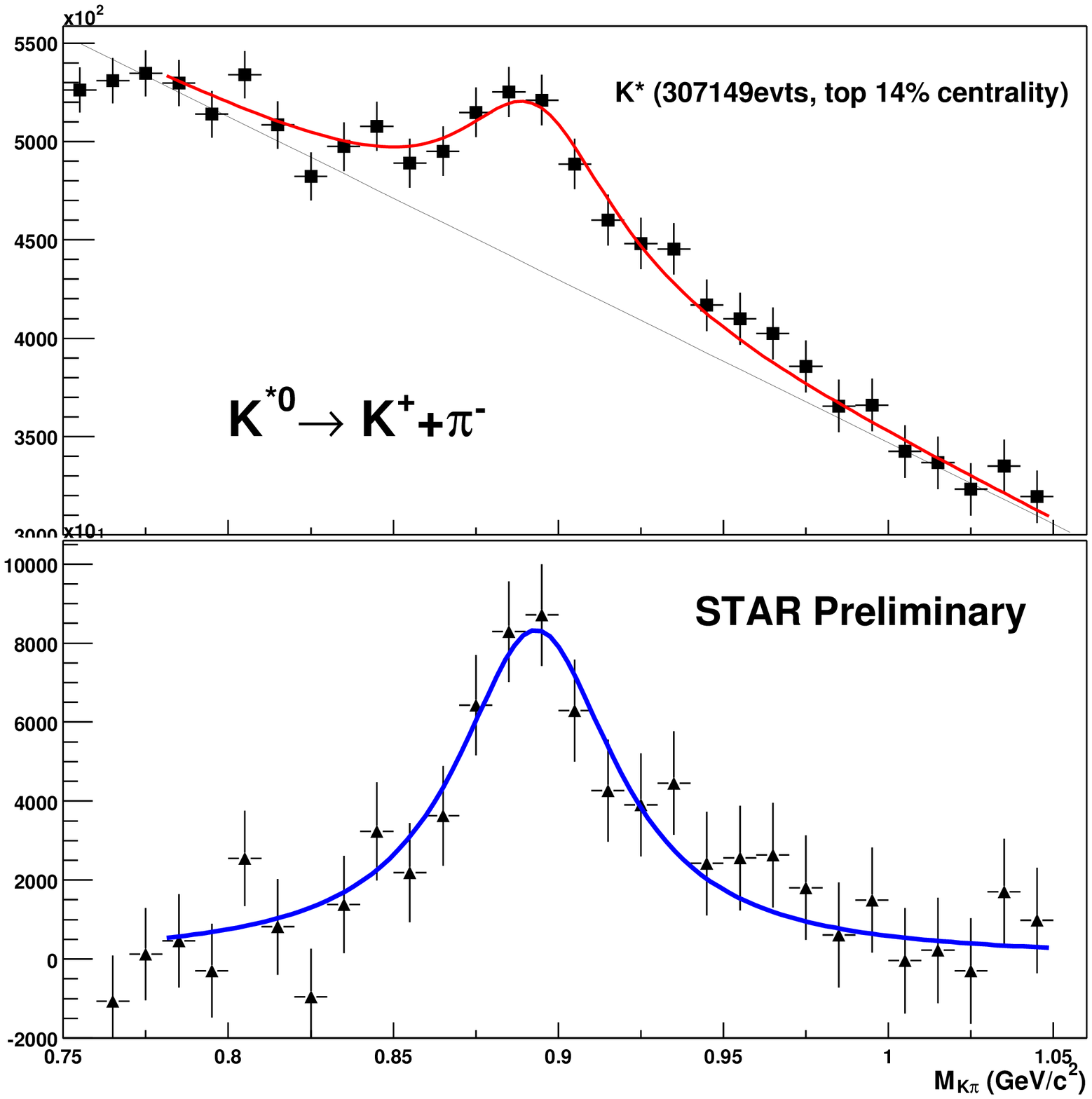,height=2.6in}
\vspace{-0.45in}
\caption{\protect{$K^{\star0}$} mass plot.\vspace{-0.7cm}
}
\label{fig:kstar}
\end{minipage}
\begin{minipage}[b]{.46\linewidth}
\centering
\epsfig{file=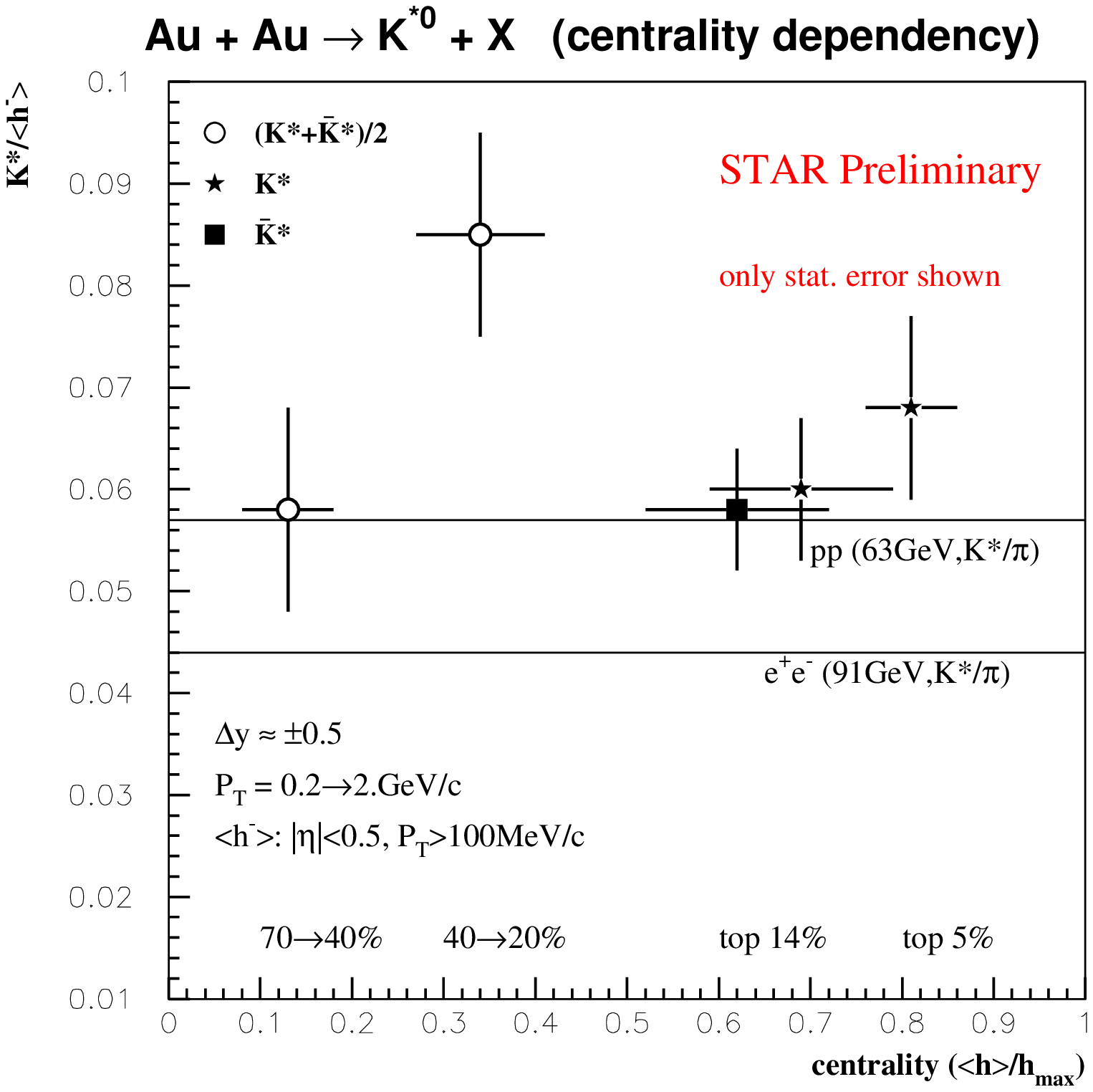,height=2.7in}
\vspace{-0.45in}
\caption{\protect{$K^{\star0}/h-$} for four centrality bins.\vspace{-0.7cm}
}
\label{fig:kpi}
\end{minipage}
\hspace{\fill}
\vspace{-0.5cm}
\end{figure}

$K^{\star0}\rightarrow K^{+}\pi^{-}$ and its antiparticle 
$\overline{K^{\star0}}\rightarrow K^{-}\pi^{+}$ (B.R. 67\%) can be 
reconstructed from the charged kaons and pions. 
The invariant mass of every $K\pi$ pair from the same event and a pool of 
mixed events is calculated and entered in the same-event spectra and 
the mixed-event spectra seperately. 
Usually three or more events are ``mixed'' with each other in same 
centrality bin and close in vertex position ($|\Delta Z|<10cm$). 
From the 300K 14\% central events, there are more than $14\times10^{9}$ 
pairs of the selected kaons and pions. Substantial computation is 
required to calculate both same-event spectra and mixed-event spectra. 
The invariant mass distribution of $K^{+}\pi^{-}$ after background subtraction 
is shown in Fig.\ref{fig:kstar} for 300K 14\% central events. 
The signal-to-background ratio before background subtraction is about 1/1000 
for central events, 1/50 for peripheral Au+Au interactions and 1/4 for 
pp at ISR~\cite{isr}. 
However, the important parameter of the significance of the signal 
is not S/N but the ratio of signal to the fluctuation in the background 
$S/\sqrt{N}$. We observe a $10\sigma$ $K^{\star0}$ signal above the background 
fluctuation and the raw count of $K^{\star0}$ is 2.7 per central event. 
Similarly, a $9\sigma$ $\overline{K^{\star0}}$ signal is observed above the 
$K^{-}\pi^{+}$ combinatoric background with the raw count of 2.6 per central 
event. 
The masses and widths are measured to be 
$m_{K^{\star0}}=0.893\pm0.003$ $GeV/c^{2}$, 
 $\Gamma_{K^{\star0}}=0.058\pm0.015$ $GeV/c^{2}$, 
$m_{\overline{K^{\star0}}}=0.896\pm0.004$ $GeV/c^{2}$ and 
$\Gamma_{\overline{K^{\star0}}}=0.063\pm0.011$ $GeV/c^{2}$. 
These are consistent with the 
values of $m_{K^{\star0}}=0.896$ $GeV/c^{2}$ and 
$\Gamma_{K^{\star0}}=0.0505$ $GeV/c^{2}$ in the 
Particle Data Book~\cite{pdg}. 
Linear and exponential shapes are used to fit the 
residual background under the signal as shown in Fig.\ref{fig:kstar}. 
The difference is about 20\% which is 
used to estimate the systematic error of the measurements. In order to get 
the yields, we have to calculate the detector acceptance and efficiency. 
This is done by embedding GEANT simulated kaons and pions to the real events 
which then go through the full reconstruction chain~\cite{manuel}. 
The overall acceptance and efficiency factor $\epsilon$ depends on 
centrality, $p_{T}$ and rapidity of the parent 
and daughter particles. $\epsilon$ changes from under 20\% for $p_{T}\simeq0$ 
to above 50\% for $p_{T}\simeq2.0$ GeV/$c$. In current studies, there is only 
one integrated $p_{T}$ and rapidity bin at each centrality from the data 
analysis. An inverse slope of $T=300MeV$ and flat in rapidity are assumed for 
$K^{\star0}$ to calculate the correction factor $\epsilon$. $\epsilon=31\%$ 
for central events and increases for lower multiplicity events. However, 
because the efficiency increases with $p_{T}$, an assumed $T=600MeV$ results 
in $\epsilon>40\%$. 
\section{Results and Discussion}
\vspace{-0.4cm}
We take $K^{\star0}/h^{-}$ yield ratio and compare the results for different 
centralities and with those from pp at ISR~\cite{isr} and $e^{+}e^{-}$ at 
LEP~\cite{lep}. $h^{-}$ is the corrected total primary 
negatively charged hadrons with $|\eta|<0.5$ and 
$0.1<p_{T}<2.0$ GeV/$c$~\cite{manuel}. 
The preliminary results as shown in Fig.\ref{fig:kpi} are 
${{(K^{\star0}+\overline{K^{\star0}})}}/2h^{-}=0.058\pm0.01$ 
($70\%\rightarrow40\%$), 
${{(K^{\star0}+\overline{K^{\star0}})}}/2h^{-}=0.085\pm0.01$ 
($40\%\rightarrow20\%$), 
$\overline{K^{\star0}}/h^{-}=0.058\pm0.006$, 
$K^{\star0}/h^{-}=0.06\pm0.007$ (top 14\%)  
and $K^{\star0}/h^{-}=0.068\pm0.009$ (top 5\%) for four centrality bins. 
Errors are statistical only and the systematic errors are 
estimated to be about 25\%. 
These results are to be compared with $K^{\star0}/\pi=0.044\pm0.003$ from 
$e^{+}e^{-}$ at $\sqrt{s}=91$ GeV and $K^{\star0}/\pi=0.057\pm0.009\pm0.009$  
from pp at $\sqrt{s}=63$ GeV. We observe that the $K^{\star0}/h^{-}$ does not 
change much from low multiplicity to high multiplicity and is 
compatible to $K^{\star0}/\pi$ in elementary particle collisions. It is 
noticed that the composition of $h^{-}$ has on the order of 80\% $\pi^{-}$ and 
the $h^{-}$ per participant pair in central Au+Au has an increase of 30\% 
from $p\overline{p}$ collision at same energies~\cite{manuel}. This implies 
about 50\% increase of the $K^{\star0}$ per participant pair. 
A better comparison could be made for $K^{\star0}/K$ since they have same 
quark content and only differ in spins. By simple spin counting, the vector 
meson to meson (pseudoscalar+vector) ratio is $V/(P+V)=0.75$. However, due to 
the mass difference between these two states, $V/(P+V)$ is much smaller from 
elementary collisions. Preliminary comparison indicates that the 
$K^{\star0}/K$ of $0.42\pm0.14$ in central Au+Au at RHIC is between the 
results of $K^{\star0}/K^{\pm}=0.64(0.55)\pm0.09\pm0.03$ at ISR and 
$K^{\star0}/K=0.32\pm0.02$ at LEP. The preliminary kaon results are from
 ~\cite{manuel}.

However, since the lifetime of $K^{\star0}$ is short ($c\tau=4fm$) and 
is comparable to the time scale of the evolution of the system, we need to 
take into account the surviving possibility of $K^{\star0}$ when comparing 
the results from Au+Au with those from pp. For example, during the time 
$\Delta t$ between chemical freeze-out and kinetic freeze-out, the daughter 
$K\pi$ from $K^{\star0}$ decay may rescatter and the $K^{\star0}$ may not be 
reconstructed. A simple model simulation which assumes that a $K^{\star0}$ 
decaying before kinetic freeze out can not be reconstructed experimentally 
shows that $\Delta t$ can only be on the order of a few fm with the current 
measured $K^{\star0}/K$ ratio. The consistency of the measured masses and 
widths of $K^{\star0}$ to the PDG values seems to 
indicate that there is no small angle elastic scattering undergone by 
daughter kaon and pion when they emergy from the dense matter. 
In reality, the surviving possibility depends on $\Delta t$, 
source size and $p_{T}$ of $K^{\star0}$. This may give us a unique tool to 
measure the time of the evolution of the system.  

In conclusion, we observe $\phi(1020),K^{\star0}(892), 
\overline{K^{\star0}}(892)$ from first-year data at RHIC taken with the STAR 
TPC. The measured $K^{\star0}/h^{-}$ are compared with $K^{\star0}/\pi$ from 
pp at ISR and $e^{+}e^{-}$ at LEP. Although data indicate slight
enhancement of $K^{\star0}$ production per participant pair from pp to Au+Au, 
$K^{\star0}/K$ seems to decrease from pp to Au+Au. A more detailed study 
including momentum distributions should be done with improved statistical and 
systematic precision from future Au+Au and pp runs at RHIC. 
We plan to measure the $p_{T}$ spectra of $K^{\star0}$ and $\phi$ from 
this data set and future data sets. We will explore the feasibility of 
measuring many other resonances. 

\end{document}